\documentstyle[preprint,aps]{revtex}

\begin{document}

\preprint{SCACOU37.TEX}
\title{Solution of the 
Scalar Coulomb Bethe-Salpeter Equation}
\author{John H. Connell\footnote{connell@stcc.mass.edu}}
\address{Springfield Technical Community 
College, Springfield, Massachusetts 01105}

\date{\today}

\maketitle

\begin{abstract}

A relativistic two-body wave equation, local in 
configuration space, is derived from the  Bethe-Salpeter 
equation for two scalar particles bound by a scalar 
Coulomb interaction.  The two-body bound-state wave 
equation is solved analytically, giving a two-body 
Bohr-Sommerfeld formula whose energies agree with the 
Bethe-Salpeter equation to order $\alpha^4$ for all quantum 
states.  From the Bohr-Sommerfeld formula, along with 
the expectation values of two remaining small 
corrections, the energy levels of the scalar Coulomb 
Bethe-Salpeter equation are worked out to order $\alpha^6$ for 
all states.  

\end{abstract}
\pacs{Draft for private circulation only}

\narrowtext

\section{Introduction}
\label{intro}

For two-body atomic systems such as the hydrogen atom 
and positronium, it would be a great advantage to have 
the energy levels to order $\alpha^4$ given by a relativistic 
two-body wave equation local in configuration space and 
soluble analytically.  We would like to have a two-body 
counterpart to the one-body Coulomb Dirac equation 
\cite{BSBOOK} with its exact solution as found by Gordon 
and by Darwin in 1928.  

This is a very old problem.  In this paper we solve it 
for a system that is analogous to real atoms but 
simpler:  two scalar particles of masses $m$ and $M$, bound 
by a scalar Coulomb potential $-\alpha /r$.  

The approach is to start from the Bethe-Salpeter 
equation \cite{BSBOOK} for the system.  A 
Bethe-Salpeter equation is used because in QED it is the 
standard two-body bound-state equation, known to be 
true, and in which higher-order corrections are well 
understood.  In QED the binding interaction in the 
Coulomb gauge kernel is $-\alpha /r$.  

In the present paper we will treat the scalar Coulomb 
system's Bethe-Salpeter equation as the true equation.  
We will require that the energy levels of any reduction 
of the Bethe-Salpeter equation agree with the levels of 
the original Bethe-Salpeter equation.  

We reduce the Bethe-Salpeter equation to a two-body 
relativistic bound-state wave equation which is {\em local\/} in 
configuration space.  The locality of this reduction is in 
contrast to the Salpeter reduction \cite{BSBOOK} which 
has been the standard for the last half century.  The 
Salpeter reduction contains operators such as $\sqrt {-\nabla^2+m^
2}$ 
which are non-local in configuration space and difficult 
to calculate with.  

The relativistic two-body Coulomb wave equation has an 
analytic solution. The energy levels are expressed by a 
two-body Bohr-Sommerfeld formula.

The Bohr-Sommerfeld formula predicts the correct 
energy levels to order $\alpha^4$.  They agree to order $\alpha^4$ 
with the Bethe-Salpeter energy levels.  

The formalism also allows calculation of the 
Bethe-Salpeter energy levels to order $\alpha^5$ and $\alpha^6$ .  
Because the levels up to first relativistic order, $\alpha^4$, are 
given by an exact solution, only first-order perturbation 
theory is needed for the level calculations up to order 
$\alpha^6$.  This is in contrast to the standard Salpeter 
reduction, in which the first relativistic order ($\alpha^4$) 
already needs first-order perturbation theory, while 
higher orders would need second-order perturbation 
theory.  In the present paper, explicit expressions for 
the Bethe-Salpeter equation's energy levels up to order 
$\alpha^6$ are given for every quantum state.  

In Section II of this paper the Bethe-Salpeter equation 
for the scalar system is written down, its Salpeter 
reduction is reviewed, and its energy levels to first 
relativistic order ($\alpha^4$) are calculated from standard 
perturbation theory for later comparison.

In Section III the local relativistic two-body wave 
equation is derived from the Bethe-Salpeter equation.  It 
is proven that the wave equation's first-order 
relativistic corrections are exactly the same as those of 
the originating Bethe-Salpeter equation for any static 
scalar interaction, so that to calculate relativistic 
corrections in first-order perturbation theory it is just 
as accurate to use the relativistic two-body wave 
equation as the Salpeter reduction.  

In Section IV, the relativistic two-body Coulomb wave 
equation is solved analytically and a two-body 
Bohr-Sommerfeld formula for the energy levels is 
derived which does predict the energy levels of the 
Bethe-Salpeter equation correctly to order $\alpha^4$.  

In Section V the formalism is used to calculate the 
Bethe-Salpeter energy levels up to order $\alpha^6$.  One set of 
corrections is given by a simple expansion of the 
Bohr-Sommerfeld formula to order $\alpha^6$.  The other two 
correction terms are calculated in first-order 
perturbation theory.  

To our knowledge the work here is the first analytic 
solution of a relativistic reduction of a Coulomb 
Bethe-Salpeter equation accurate to order $\alpha^4$.  It may 
also contain the first accurate two-body 
Bohr-Sommerfeld formula.  We have also demonstrated 
that corrections to the energy levels of a Coulomb 
Bethe-Salpeter equation up to order $\alpha^6$ can be calculated 
in first-order perturbation theory, once the 
solution is found analytically to order $\alpha^4$.  

It is hoped that techniques like these may one day lead 
to similar results for real atomic systems such as 
positronium and the hydrogen atom.

\section{Scalar Bethe-Salpeter equation and Salpeter reduction} 
\subsection{Bethe-Salpeter Equation}

The notation will be the following.  Let the 
masses of the bound particles be $m$,$M$.  Let the mass of 
the bound state be denoted by $E$, while defining the 
bound-state wave number below threshold as $\beta$, as well 
as the particles' individual CM bound-state energies as $t$ and 
$T$, in the following way:  
\begin{equation}E=\sqrt {m^2-\beta^2}+\sqrt {M^2-\beta^2}\label{E}\end{equation}
\begin{equation}t\equiv\sqrt {m^2-\beta^2}=\frac {E^2+(m^2-M^2)}{
2E},\label{t}\end{equation}
\begin{equation}T\equiv\sqrt {M^2-\beta^2}=\frac {E^2+(M^2-m^2)}{
2E}\label{T}\end{equation}
The CM energy-momenta of the particles are written 
$({\bf p},t+p^0)$, $(-{\bf p},T-p^0)$.  

The Bethe-Salpeter equation for the bound-state vertex 
function $\Gamma$ will be 
\begin{equation}\Gamma =I*S~\Gamma\label{VFNBSE}\end{equation}
with
\begin{equation}S({\bf p},p^0)\equiv -\:\frac {2m}{{\bf p}^2+m^2-
i\epsilon -(t+p^0)^2}\:\frac {2M}{{\bf p}^2+M^2-i\epsilon -(T-p^0
)^2}\label{S}\end{equation}
The scalar interaction kernel is written $4mMI({\bf k}^2)$, or 
$4mMI(r)$ in configuration space.  The star denotes an 
integration over 4-momentum with a factor $1/i$ present.  
The factors $2m$,$2M$ have been moved into the numerators 
of the propagators of the scalar constituent particles in 
order to make the dimensions of the scalar functions the 
same as in spin-$\frac 12$ systems.  

\subsection{Salpeter Reduction}

Defining $\phi ({\bf p})=\int dp^0S({\bf p},p^0)\Gamma ({\bf p},p^
0)/2\pi i$, equation  
(\ref{VFNBSE}) when multiplied by $S$ and integrated over 
$p^0$ gives the non-local Salpeter equation equivalent to 
(\ref{VFNBSE}): 
\begin{equation}\left[\sqrt {{\bf p}^2+m^2}+\sqrt {{\bf p}^2+M^2}
+Z({\bf p})I(r)\right]\phi ({\bf r})=E\phi ({\bf r})\label{SAL}\end{equation}
in which $Z({\bf p})$ is a correction operator on the interaction $
I(r)$:
\begin{equation}Z({\bf p})=\frac {2mM\left[\sqrt {{\bf p}^2+m^2}+\sqrt {
{\bf p}^2+M^2}\right]}{\sqrt {{\bf p}^2+m^2}\sqrt {{\bf p}^2+M^2}\left
[\sqrt {{\bf p}^2+m^2}+\sqrt {{\bf p}^2+M^2}+E\right]}\label{ZSAL}\end{equation}

\subsection{First-order Relativistic Correction to Energy 
Levels} 

To calculate the first-order relativistic corrections to 
the Schr\"odinger energy levels predicted by equation 
(\ref{SAL}) the non-relativistic Schr\"odinger equation is as 
usual defined to be 
\begin{equation}\left[m+M+\frac {{\bf p}^2}{2\mu}+I(r)\right]\phi_
0({\bf r})=E_0\phi_0({\bf r})\label{SCH}\end{equation}
in which $\mu$ is the reduced mass. The non-relativistic 
bound-state wave number $\beta_0$ will be defined by 
\begin{equation}\mbox{\rm $E_0=m+M-\beta_0^2/2\mu$}\label{E0}\end{equation}

Then the Salpeter reduction (\ref{SAL}) is expanded in 
powers of ${\bf p}^2$ and $\beta_0^2$ in the standard way, giving the 
 correction $\Delta E=E-E_0$:
\begin{equation}\Delta E=-\frac {\beta_0^4}8\left(\frac 1{m^3}+\frac 
1{M^3}\right)+\frac {\beta_0^2}{2mM}\left<\phi_0\right|I\left|\phi_
0\right>+\frac 1{2\mu}\left(\frac {\mu^2}{m^2}+\frac {\mu^2}{M^2}\right
)\left<\phi_0\right|I^2\left|\phi_0\right>\label{DELE}\end{equation}
This constitutes the first-order relativistic correction 
for the energy level of the original Bethe-Salpeter 
equation (\ref{VFNBSE}).  

In the scalar Coulomb case
\begin{equation}I(r)=-\frac {\alpha}r\label{COULPOT}\end{equation}
equation (\ref{DELE}) shows that the energy levels of 
the scalar Coulomb Bethe-Salpeter equation have this 
 correction to order $\alpha^4$:
\begin{equation}\Delta E^{(4)}_{\mbox{\rm Coul}}=\frac {\alpha^4\mu}{
N^3(2L+1)}\left(\frac {\mu^2}{m^2}+\frac {\mu^2}{M^2}\right)-\frac {
\alpha^4\mu}{8N^4}\left(1+\frac {\mu^2}{mM}\right)\label{DELECOUL}\end{equation}
where $L$ is the angular momentum and $N$ is the Bohr 
quantum number.

\section{Relativistic Wave Equation}

\subsection{Derivation}

We return to the original Bethe-Salpeter equation 
(\ref{VFNBSE}) and reduce it to a three-dimensional form 
in a manner different from that of Salpeter.

Because $S$ peaks sharply at $p^0=0$ when ${\bf p}^2$ is not too 
large, it can be approximated \cite{JHC67,TOD71,JHC91} by
\begin{equation}S_0({\bf p},p^0)\equiv -\frac {4mM}{2E}\frac {2\pi 
i\delta (p^0)}{({\bf p}^2+\beta^2)}\label{S0}\end{equation}
The error of the approximation is defined by
\begin{equation}R\equiv S-S_0\label{R}\end{equation}
The Blankenbecler-Sugar \cite{BLS} correction interaction 
$U$ is defined by the equation
\begin{equation}U=I+I*R~U\label{BLSPOT}\end{equation}
Then from the Bethe-Salpeter equation (\ref{VFNBSE}) it 
is easy to deduce the equation $\Gamma =U*S_0~\Gamma$.  Next, 
defining $\phi ({\bf p})$ by $S_0~\Gamma =2\pi i\delta (p^0)\phi 
({\bf p})$, it follows that, 
with the notation $U=U(p;q)$, 
\begin{equation}({\bf p}^2+\beta^2)\phi ({\bf p})=-\frac {4mM}{2E}
\int\frac {d\,{\bf q}}{(2\pi )^3}U({\bf p},0;{\bf q},0)\phi ({\bf q}
)\label{MYEQN}\end{equation}
Equation (\ref{MYEQN}) is exact, in terms of the original 
Bethe-Salpeter equation (\ref{VFNBSE}).  

For a static kernel $I({\bf k}^2)$,  equation (\ref{BLSPOT}) 
for $U$ simplifies, because in its solution by iteration the 
integration over the relative-energy variable only acts 
on $R$.  After that integration, equation (\ref{BLSPOT}) 
becomes 
\begin{equation}U({\bf p},0;{\bf q},0)=I[({\bf p}-{\bf q})^2]+\int\frac {
d\,{\bf l}}{(2\pi )^3}I[({\bf p}-{\bf l})^2]R({\bf l}^2)U({\bf l}
,0;{\bf q},0)\label{UEQN}\end{equation}
in which
\begin{equation}R({\bf l}^2)=\frac {mM}E\left\{\frac 1{\sqrt {{\bf l}^
2+m^2}\left(\sqrt {{\bf l}^2+m^2}+t\right)}+\frac 1{\sqrt {{\bf l}^
2+M^2}\left(\sqrt {{\bf l}^2+M^2}+T)\right)}\right\}\label{Rint}\end{equation}
where the energy-dependent constants $t$ and $T$ were 
defined in (\ref{t}) and (\ref{T}).

\subsection{Local Equation}

Equation (\ref{UEQN}) for $U$ leads immediately to a local 
wave equation accurate to first relativistic order.

Because in many systems the momentum is small 
compared to the constituent masses, it is a reasonable 
starting point to approximate (\ref{Rint}) by its value at 
${\bf l}^2=0${\bf :  }
\begin{equation}R_E\equiv R(0)=\frac {mM}E\left[\frac 1{m(m+t)}+\frac 
1{M(M+T)}\right]\label{RE}\end{equation}
Then in configuration space equation (\ref{UEQN}) can be 
iterated once to give a local approximation to $U$:  
\begin{equation}U_E(r)=I(r)+R_EI^2(r)\label{UE4}\end{equation}
Defining in addition the quantity
\begin{equation}Z_E\equiv\frac {m+M}E\label{ZE}\end{equation}
equation (\ref{MYEQN}) gives
\begin{equation}\left\{{\bf p}^2+2\mu Z_E\left[I(r)+R_EI^2(r)\right
]\right\}\phi ({\bf r})=-\beta^2\phi ({\bf r})\label{LWEQN}\end{equation}
which is a two-body wave equation local in configuration 
space.  Equation (\ref{LWEQN}) contains energy-dependent 
constants, instead of the non-local operators of the 
Salpeter reduction (\ref{SAL}).  The eigenvalue $\beta^2$ 
determines the energy $E$  through equation (\ref{E}).  

\subsection{Equivalence Proof}

Equation (\ref{LWEQN}) can be expanded around the 
Schr\"odinger equation (\ref{SCH}) to calculate the  energy 
corrections given by (\ref{LWEQN}) to the first 
relativistic order.  

Expression (\ref{E}) for the energy must be expanded 
to fourth order in $\beta_0$, so that 
\begin{equation}\Delta E=-\frac {\beta_0^4}8\left(\frac 1{m^3}+\frac 
1{M^3}\right)-\frac 1{2\mu}\Delta\beta^2\label{LODELE}\end{equation}
In evaluating $\Delta\beta^2\equiv\beta^2-\beta_0^2$ from (\ref{LWEQN}), it is only 
necessary to express $R_E$ to zero order in $\beta_0$, as 
$(\mu^2/m^2+\mu^2/M^2)/2\mu$.  In addition $Z_E$ need only be 
expanded to second order in $\beta_0$, as $1+\beta_0^2/2mM$; in 
the $I^2$ term $Z_E$ can be taken to be $1$.

From these values (\ref{LWEQN}) immediately gives the 
first-order correction 
\begin{equation}\Delta\beta^2=-\frac {\beta_0^2}{m+M}\langle I\rangle 
-\left(\frac {\mu^2}{m^2}+\frac {\mu^2}{M^2}\right)\langle I^2\rangle\label{DELBSQ}\end{equation}
Substitution of (\ref{DELBSQ}) into (\ref{LODELE}) shows  
that the expression for $\Delta E$ is the same as the 
expression (\ref{DELE}) derived by means of the Salpeter 
reduction.  

This constitutes a proof that the local relativistic 
two-body wave equation (\ref{LWEQN}) correctly gives 
the first-order relativistic energy corrections to the 
Bethe-Salpeter equation (\ref{VFNBSE}).  A similar proof 
for spin-$\frac 12$ particles was derived some time ago 
\cite{JHCSLAC}.

\subsection{Infinite-Mass Limit}

In the limit $M\to\infty$, equation (\ref{LWEQN}) becomes 
\begin{equation}\left\{{\bf p}^2+2mI(r)+\frac {2m}{m+t}I^2(r)\right
\}\phi ({\bf r})=-\left(m^2-t^2\right)\phi ({\bf r})\label{MINF}\end{equation}
where $t$, defined in (\ref{t}), is the bound-state energy of 
the lighter particle.

To first relativistic order the term $2mI^2/(m+t)$ can be 
replaced by $I^2$.  Then (\ref{MINF}) becomes 
\begin{equation}\left\{{\bf p}^2+\left[m+I(r)\right]^2\right\}\phi 
({\bf r})=t^2\phi ({\bf r})\label{KG}\end{equation}
which is the customary Klein-Gordon equation for a 
scalar particle in a fixed field.

\section{Scalar Coulomb Solution}

Specialising to the scalar Coulomb interaction 
(\ref{COULPOT}), the Blankenbecler-Sugar correction 
potential given by (\ref{UEQN}) becomes to second order 
\begin{equation}U_{\mbox{\rm Coul}}(r)=-\frac {\alpha}r+R_E\frac {
\alpha^2}{r^2}\label{UECOUL}\end{equation}
and the relativistic two-body wave equation 
(\ref{LWEQN}) becomes 
\begin{equation}\left\{{\bf p}^2+2\mu Z_E\left[-\frac {\alpha}r+R_
E\frac {\alpha^2}{r^2}\right]\right\}\phi =-\beta^2\phi\label{COULWVEQN}\end{equation}

The radial reduction of equation (\ref{COULWVEQN}) has 
the same singularities as that of the Coulomb 
Schr\"odinger equation, and can be solved the same way.  
For angular momentum $L$ the radial component can be 
expanded for radial quantum numbers $n=0,1,2,\cdots$ as 
\begin{equation}e^{-\beta r}(\beta r)^{\epsilon}\sum^n_{i=0}a_i(\beta 
r)^i\label{SERIES}\end{equation}
which as usual  gives a recurrence relation between the 
coefficients $a_i$. The existence of $a_0$, and the termination 
of the series, require the conditions
\begin{equation}\epsilon =\sqrt {\left(L+\frac 12\right)^2+\alpha^
22\mu Z_ER_E}-\frac 12\label{EPS}\end{equation}
and
\begin{equation}\beta =\frac {Z_E\alpha\mu}{n+1+\epsilon}\label{BETA}\end{equation}

Then from (\ref{BETA}) and  expression (\ref{E}) for 
the energy $E$, it is easy to deduce the Bohr-Sommerfeld 
formula
\begin{equation}E=\sqrt {m^2+M^2+2mM\sqrt {1-\frac {\alpha^2}{(n+
1+\epsilon )^2}}}\label{BOHRSOMM}\end{equation}
Expanding in the usual way, with $Z_E$ and $R_E$ inside $\epsilon$ 
only needed to zero order, it is found that the 
Bohr-Sommerfeld expression (\ref{BOHRSOMM}) does 
predict the correct $\alpha^4$ correction (\ref{DELECOUL}) to 
the energy levels of the scalar Coulomb Bethe-Salpeter 
equation, as expected.

\[\]

\section{Energy Levels to Order $\alpha^6$}

\subsection{Introduction}

Up to this point it has only been shown that the 
reduction of the Bethe-Salpeter equation to a relativistic 
two-body wave equation as described in the previous 
section reproduces the results of the standard Salpeter 
reduction to order $\alpha^4$, albeit through an analytic solution 
instead of perturbation theory.  One way of investigating 
whether the present reduction may be more useful than 
the method of Salpeter would be to find out whether 
higher-order terms in the energy levels can be calculated 
more easily than the Salpeter reduction allows.  In the 
present section we will calculate the energy levels of 
the Bethe-Salpeter equation (\ref{VFNBSE}) analytically 
to order $\alpha^6$ for all quantum states, using only 
first-order perturbation theory.

With the definition
\begin{equation}\Delta R({\bf l}^2)\equiv R({\bf l}^2)-R_E\label{DELR}\end{equation}
the expansion of equation  (\ref{UEQN}) for $U$ becomes 
(with $I$ standing for $-\alpha /r$)
\begin{equation}U=U_{\mbox{\rm Coul}}+I\,\Delta R\,I+IRIRI+IRIRIR
I+\cdots\label{UEXPN}\end{equation}

We will evaluate the contributions to the energy levels 
up to order $\alpha^6$ due to:
\begin{itemize}
\item\  $U_{\mbox{\rm Coul}}$, whose levels are contained in the 
Bohr-Sommerfeld formula (\ref{BOHRSOMM});
\item\   $I\,\Delta R\,I$, which will have an $\alpha^5$ contribution for 
$L=0$, as well as $\alpha^6$ contributions;
\item\   the order-$\alpha^6$ contribution of the rest of the 
series,
\begin{equation}U_6=IRIRI+IRIRIRI+\cdots\label{U6}\end{equation}
\end{itemize}

\[\]

\subsection{Bohr-Sommerfeld Formula}

The Bohr-Sommerfeld expression (\ref{BOHRSOMM}) can 
be expanded further to get the sixth-order contribution 
to the energy of the Blankenbecler-Sugar potential $U_{\mbox{\rm Coul}}$, equation 
(\ref{UECOUL}). Defining the dimensionless constant
\begin{equation}c\equiv\frac {\mu^2}{m^2}+\frac {\mu^2}{M^2}
\label{c}\end{equation}
the sixth-order energy contribution is
\begin{eqnarray}
\Delta E^{(6)}_{\mbox{\rm Bohr-Somm}}&=&\alpha^6\mu\left\{\frac {
4c-7-c^2}{64N^6}+\right.\nonumber\\
&&\left.+\frac {12c-5c^2-1}{8N^5(2L+1)}-\frac {3c^2}{2N^4(2L+1)^2}
-\frac {c^2}{N^3(2L+1)^3}\right\}\label{DELE6BS}\end{eqnarray}
We recall that $L$ is the angular momentum and $N$  is the 
Bohr quantum number. To obtain this expression the 
constants $Z_E$, $R_E$ and $\epsilon$ are expanded to the  
order in $\alpha$ needed.

\subsection{Correction $I\,\Delta R\,I$}

Following the steps outlined in Subsection A above, we 
calculate the $\alpha^5$ and $\alpha^6$ energy corrections due to the 
term $I\,\Delta R\,I$ in (\ref{UEXPN}), where $\Delta R$ is defined in 
(\ref{DELR}). The correction is
\begin{equation}\Delta E^{(6)}_{\Delta R}=\left<\phi\right|I\,\Delta 
R\,I\left|\phi\right>\label{DELE6DELREXP}\end{equation}
Up to order $\alpha^6$ it is sufficient to use non-relativistic 
wavefunctions.  Use of the Schr\"odinger equation 
(\ref{SCH}) then gives 
\begin{equation}\Delta E^{(6)}_{\Delta R}=\frac 1{(2\mu )^2}\int\frac {
d\,{\bf p}}{(2\pi )^3}\phi^{*}_0({\bf p})\left[{\bf p}^2+\beta_0^
2\right]^2\Delta R({\bf p}^2)\phi_0({\bf p})\label{DELE6DELR}\end{equation}

To evaluate the integral (\ref{DELE6DELR}) for every 
quantum state, it is convenient to change co\"ordinates 
from {\bf p}-space to the surface of Schwinger's unit sphere 
\cite{SCHW64} in Cutkosky's 4-space \cite{JHC91}.  The 
transformation is (with $i=1,2,3$) 
\begin{equation}\xi_i=\frac {2\beta {\bf p}_i}{{\bf p}^2+\beta^2}
,~~~~\xi_4=\frac {{\bf p}^2-\beta^2}{{\bf p}^2+\beta^2}\label{SCHWXI}\end{equation}
(Henceforth $\beta$ is written for $\beta_0$.)  The polar co\"ordinates 
on the unit sphere are conventionally denoted by 
$(\theta_1,\theta_2,\phi )$ \cite{ERD} where $\theta_2$, $\phi$ are the usual angles of 
3-space, and 
\begin{equation}x\equiv\cos\theta_1=\xi_4\label{x}\end{equation}
The element of surface area is \cite{ERD} 
\begin{equation}d\Omega =(\sin\theta_1)^2(\sin\theta_2)d\theta_1d
\theta_2d\phi\label{DOMEGA}\end{equation}
and $d\Omega$ and $d{\bf p}$ are related by \cite{SCHW64} 
\begin{equation}d\Omega =\left[\frac {2\beta}{{\bf p}^2+\beta^2}\right
]^3d{\bf p}\label{DREL}\end{equation}

Surface harmonics on Schwinger's unit sphere are 
denoted by $Y_{NLM}$, where $N-1\ge L\ge |M|\ge 0$.  The 
quantum numbers $L,M$ have their usual meaning in 
3-space, and $N=1,2,3,\ldots$ is the Bohr quantum number.  
As usual they are related by $N=n+L+1$ where 
$n=0,1,2,\ldots$ is the radial quantum number in ${\bf p}$-space.

The standard representation of the normalised surface 
harmonics is \cite{ERD} 
\begin{equation}Y_{NLM}(\theta_1,\theta_2,\phi )=2^{L+1}\Gamma (L
+1)\sqrt {\frac {n!\,N}{2\pi\Gamma (N+L+1)}}\,(1-x^2)^{\frac L2}C^{
L+1}_n(x)Y_{LM}(\theta_2,\phi )\label{Y}\end{equation}
in which $Y_{LM}(\theta_2,\phi )$ is normalised on the surface of the 
unit sphere in 3 dimensions, and the coefficient is 
determined by the orthonormality relation \cite{ERD} 
\begin{equation}\int^{+1}_{-1}(1-x^2)^{L+\frac 12}C^{L+1}_m(x)C^{
L+1}_n(x)\,dx=\delta_{m,n}\frac {\pi\,\Gamma (N+L+1)}{n!\,N\left[
\Gamma (L+1)\right]^22^{2L+1}}\label{CNORM}\end{equation}

The momentum-space eigenfunctions $\phi_0({\bf p})$ are 
\cite{SCHW64} proportional to $(1-x)^2Y_{NLM}$.  With the 
normalisation requirement 
\begin{equation}\int\frac {d\,{\bf p}}{(2\pi )^3}\left|\phi_0({\bf p}
)\right|^2=1\label{PHINORM}\end{equation}
we find that
\begin{equation}\phi_0({\bf p})=\left[\frac {2\pi}{\beta}\right]^{\frac 
32}(1-x)^2\,Y_{NLM}(\theta_1,\theta_2,\phi )\label{PHI0}\end{equation}

Then substitution of equation (\ref{PHI0}) into 
(\ref{DELE6DELR}) gives
\begin{equation}\Delta E^{(6)}_{\Delta R}=\frac {2^{2L+1}n!N\left
[\Gamma (L+1)\right]^2}{\pi\Gamma (N+L+1)}\frac {\beta^4}{\mu^2}I_{
L,n}\label{DELE6ILn}\end{equation}
in which
\begin{equation}I_{L,n}\equiv\int^{+1}_{-1}dx\,(1-x^2)^{L+\frac 1
2}\left[C^{L+1}_n(x)\right]^2\frac {\Delta R({\bf p}^2)}{1-x}\label{ILn}\end{equation}

Although $\Delta R(0)=0$, it is not possible to expand $\Delta R(
{\bf p}^2)$ to 
first order as a Taylor series in ${\bf p}^2$, because the 
integrals in (\ref{DELE6DELR}) and (\ref{ILn}) would 
diverge when $L=0$.  In order to evaluate $I_{L,n}$ for all 
states including $L=0$ it is necessary to express $\Delta R({\bf p}^
2)$ 
exactly.  

With successive definitions, referring to equations 
(\ref{SCHWXI}) and (\ref{x}):
\begin{equation}a\equiv\frac {2\beta^2}{t^2},~~~~~A\equiv\frac {2
\beta^2}{T^2}\label{aA}\end{equation}
\begin{equation}D(y,a,t,m)\equiv a\left\{\frac {5\sqrt y+4\sqrt {
y+a}}{3(\sqrt y+\sqrt {y+a})^2}+\frac {t^2(m+2t)}{2m(m+t)^2}\sqrt 
y\right\}\label{D}\end{equation}
\begin{equation}F(x)\equiv -\frac {mM}E\left\{\frac 1{t^2}D(1-x,a
,t,m)+\frac 1{T^2}D(1-x,A,T,M)\right\}\label{F}\end{equation}
it can be shown that
\begin{equation}\frac {\Delta R({\bf p}^{2)}}{\sqrt {1-x}}=\frac 
d{dx}F(x)\label{TOTDERIV}\end{equation}

Expression (\ref{ILn}) can now be integrated by parts.  To 
the required order, it becomes
\begin{equation}I_{L,n}=-\alpha\,\delta_{L,0}\,\frac {8N}{3\mu}\left
[\frac {\mu^3}{m^3}+\frac {\mu^3}{M^3}\right]-J_{L,n}\label{IJ}\end{equation}
in which
\begin{equation}J_{L,n}=\int^{+1}_{-1}d\,x\,F(x)\frac d{dx}\left\{\sqrt {
1+x}(1-x^2)^L\left[C^{L+1}_n(x)\right]^2\right\}\label{JLn}\end{equation}
Note that the $\delta_{L,0}$ term is of order $\alpha$, not $\alpha^
2$.  It will 
give a contribution of order $\alpha^5\mu$ to the energy.  In 
atomic physics the Coulomb potential gives an $\alpha^5$ 
contribution when $L=0$.  In the present model and 
formalism that term appears in the $\Delta R$ correction.

The remaining integral (\ref{JLn}) stays convergent as 
$a,A\to 0$ inside the square roots in the functions $D$ 
contained in $F(x)$.  To the required order that limit may 
be taken, and  (\ref{JLn}) becomes to lowest order 
\begin{equation}J_{L,n}=-\alpha^2\frac 1{\mu N^2}\left[\frac {\mu^
4}{m^4}+\frac {\mu^4}{M^4}\right]\left(K^{(1)}_{L,n}+K^{(2)}_{L,n}\right
)\label{JK}\end{equation}
in which
\begin{eqnarray}K^{(1)}_{L,n}&=&\frac 34\int^{+1}_{-1}d\,x\sqrt {1-
x}\frac d{dx}\left\{\sqrt {1+x}(1-x^2)^L\left[C^{L+1}_n(x)\right]^
2\right\}\label{KLn1}\\K^{(2)}_{L,n}&=&\frac 32\int^{+1}_{-1}d\,x\frac 1{\sqrt {
1-x}}\frac d{dx}\left\{\sqrt {1+x}(1-x^2)^L\left[C^{L+1}_n(x)\right
]^2\right\}\label{KLn2}\end{eqnarray}

The expression for $K^{(1)}_{L,n}$ can be integrated by parts for 
all $L$.  The expression for $K^{(2)}_{L,n}$ can be integrated directly 
by parts for $L\ge 1$, while for $L=0$ the integration by 
parts may be done by subtracting $\left[C^{L+1}_n(1)\right]^2$ from  $\left
[C^{L+1}_n(x)\right]^2$ 
in the integrand beforehand. The subtracted part is 
calculated separately and added back after the integral is 
evaluated. With that understanding, we have
\begin{eqnarray}
K^{(1)}_{L,n}&=&+\frac 38\int^{+1}_{-1}\frac {(1-x^2)^{L+\frac 12}\left
[C^{L+1}_n(x)\right]^2}{1-x}dx\label{KLn1PT}\\
K^{(2)}_{L,n}&=&-\frac 34\int^{+1}_{-1}\frac {(1-x^2)^{L+\frac 12}\left
[C^{L+1}_n(x)\right]^2}{(1-x)^2}dx\label{KLn2PT}\end{eqnarray}

To evaluate these $K_{L,n}$ integrals it has been necessary 
to derive the following equation \cite{RG} (in which 
$m\le n$):
\begin{eqnarray}
&&\int^{+1}_{-1}dx\frac {(1-x^2)^{L+\frac 12}C^{L+1}_m(x)C^{L+1}_
n(x)}{z-x}\nonumber\\
=&&\frac {\sqrt {\pi}\Gamma (N+L+1)}{2^{N+L}\Gamma (L+1)\Gamma (N
+\frac 12)}C^{L+1}_m(z)\int^{+1}_{-1}dt\frac {(1-t^2)^{n+L+\frac 
12}}{(z-t)^{n+1}}\label{z}\end{eqnarray}
  
With $m=n$ the limit $z\to 1$ gives
\begin{equation}K^{(1)}_{L,n}=\frac 38\frac {\sqrt {\pi}\Gamma (L
+\frac 12)\Gamma (N+L+1)}{\Gamma (n+1)\,\Gamma (L+1)\,\Gamma (2L+
2)}\label{KLn1VAL}\end{equation}

To evaluate (\ref{KLn2PT}) one can differentiate (\ref{z}) 
with respect to $z$ and take the limit $z\to 1$.  For $L=0$ the 
subtracted integral is calculated before taking the limit 
$z\to 1$; the value is then finite. For all $L$, including $L=0$, 
the result is
\begin{equation}K^{(2)}_{L,n}=\frac 34\frac {\sqrt {\pi}\Gamma (L
+\frac 12)\Gamma (N+L+1)}{\Gamma (n+1)\Gamma (L+1)\Gamma (2L+2)}\,\frac {
1-4N^2}{(2L-1)(2L+3)}\label{KLn2VAL}\end{equation}
These evaluations are carried out using standard representations and 
properties of the beta function and the hypergeometric 
function.

Working back from these results to the original 
expression (\ref{DELE6ILn}) for the energy correction, we 
finally have
\begin{eqnarray}
\Delta E^{(6)}_{\Delta R}&=&-\alpha^5\mu\delta_{L,0}\frac {16}{3\pi 
N^3}\left[\frac {\mu^3}{m^3}+\frac {\mu^3}{M^3}\right]+\nonumber\\
&&+\alpha^6\mu\frac 3{2N^5(2L+1)}\left[\frac {\mu^4}{m^4}+\frac {
\mu^4}{M^4}\right]\left[\frac 12+\frac {1-4N^2}{(2L+3)(2L-1)}\right]
\label{DELE6DELRVAL}\end{eqnarray}

The reader is reminded that this is the energy 
correction to order $\alpha^6$ due to the correction 
$\Delta R=R({\bf p}^2)-R_E$, and that its calculation used first-order 
perturbation theory only.

\subsection{Correction $IRIRI+IRIRIRI+\cdots$}

Finally we address the rest of the series for $U$, given 
by (\ref{U6}).  

It will be proven below that to the required order $R({\bf p}^2)$ 
can be approximated by its lowest-order value $c/2\mu$, 
where the dimensionless function of the masses $c$ is 
defined by (\ref{c}).  Then the series (\ref{U6}) becomes 
to lowest order
\begin{eqnarray}
U^0_6(r)&=&I(c/2\mu )I(c/2\mu )I+I(c/2\mu )I(c/2\mu )I(c/2\mu )I+
\ldots\label{U60SERIES}\\
&=&\frac {\alpha^3(c/2\mu )^2}{r^2(r+\alpha c/2\mu )}
\label{U60}\end{eqnarray}
The energy correction up to order $\alpha^6$ is given by the 
expectation value of (\ref{U60}) over non-relativistic 
Coulomb wavefunctions.

For $L\ge 1$ the lowest-order expectation value is just 
dictated by the expectation value of $1/r^3$. It is
\begin{equation}\left.\Delta E^{(6)}_{U6}\right|_{L\ge 1}=\left\langle 
U_6(r)\right\rangle_{L\ge 1}=-\alpha^6\mu\frac {c^2}{2N^3L(L+1)(2
L+1)}\label{DELEU6L1}\end{equation}

For $L=0$ a logarithm occurs.  Unfortunately expectation 
values of quantities of the form (\ref{U60}) over 
Coulomb wavefunctions do not seem to be listed in 
standard references.  By expanding the 
polynomial in the radial wavefunctions the result can 
easily be found as a double sum from $0$ to $n$.  But in 
case this kind of integral arises  later in real atomic 
two-body systems, at the cost of a few more lines of 
calculation we present an evaluation in closed form.

In terms of the variable $z=2\beta_0r=2\alpha\mu r/N$, the radial 
wavefunction is known to be $z^Le^{-z/2}F(-n,2L+2,z)$ up to 
normalisation, where $F$ is the confluent hypergeometric 
function. Recalling that $L=0$, and defining 
$a=\alpha^2c/N$, we need to evaluate
\begin{equation}K_a\equiv\int^{\infty}_0e^{-z}\frac 1{z+a}\left[F
(-n,2,z)\right]^2dz\label{Ka}\end{equation}

While this expression has not been found in any 
reference, Landau and Lifshitz \cite{LL} give an 
evaluation of a related integral:  
\begin{eqnarray}
J_{\nu}&\equiv&\int^{\infty}_0e^{-z}z^{\nu -1}\left[F(-n,\gamma ,
z)\right]^2dz\label{Jnu}\\
&=&\frac {\Gamma (\nu )n!}{(\gamma )_n}\sum^n_{i=0}\frac {(-n)_i(
\nu -\gamma +1)_i(\gamma -\nu )_i}{[i!]^2(\gamma )_i}\label{JnuVAL}
\end{eqnarray}
where as usual $(\kappa )_k=\kappa (\kappa +1)\cdots (\kappa +k-1
)$, $(\kappa )_0=1$.  Here 
the quantity $\gamma$ will be $2$ since $L=0$.  The value $\nu =3$ 
gives the normalisation integral $J_3=2$.  

In $K_a$ the parameter $a$ is small and a $\log(1/a)$ term will 
dominate, followed by $\mbox{\rm O}(1)$ terms.  These are the terms 
needed to find the energy to order $\alpha^6$.  Also, if $\nu$ is 
taken to be small in $J_{\nu}$ a $1/\nu$ term will dominate and 
the next terms will be $\mbox{\rm O}(1)$.  If 
$\left[F(-n,2,z)\right]^2$ is written as $1+B(z)$, in $B(z)$ the lowest 
power of $z$ will be one.  Then it is clear that in both $K_a$ 
and $J_{\nu}$, the leading contribution of the part $B(z)$ is of order 1.  
Furthermore, those O$(1)$ terms can be calculated by 
replacing $1/(z+a)$ by $1/z$ and $z^{\nu -1}$ by $1/z$ 
respectively.  In other words, the contribution of $B(z)$ to 
the $\mbox{\rm O}(1)$ terms of both integrals is identical.  
In addition, the contribution of the ``$1$'' in each 
integral can be calculated explcitly.

Therefore the way to evaluate the unknown integral 
(\ref{Ka}) up to $\mbox{\rm O}(1)$ for small $a$ is the following.  (i) 
From the result (\ref{JnuVAL}) subtract the explicitly 
calculated contribution to $J_{\nu}$ of the ``$1$'' in $[F]^2$.  (ii) 
Evaluate the now leading O$(1)$ part of the result, 
neglecting terms of order $\nu$.  (iii) Explicitly calculate the 
contribution of the ``$1$'' to $K_a$ to orders $\log(1/a)$ and $1$, 
neglecting terms of order $a$.  (iv) Add that to the result 
(ii), finally obtaining the $\log(1/a)$ and O$(1)$ terms of $K_a$ as 
required. The result is
\begin{equation}K_a=\log\frac 1a-(n+2){\bf C}+\frac {n(4n+1)}{2(n
+1)}-(n+1)\psi (n+1)+\mbox{\rm O}(a)\label{KaVAL}\end{equation}
and hence to order $\alpha^6$
\begin{eqnarray}
&&\left.\Delta E^{(6)}_{U6}\right|_{L=0}=\left\langle U_6(r)\right
\rangle_{L=0}=\nonumber\\
&&=\alpha^6\mu\frac {c^2}{N^3}\left\{\log\frac N{\alpha^2c}-(N+1)
{\bf C}+\frac {n(4n+1)}{2N}-N\psi (N)\right\}\label{DELEU6L0}\end{eqnarray}
in which $N=n+1$, $\psi (z)\equiv d\log[(\Gamma (z)]/dz$ and ${\bf C}$ is Euler's 
constant $0.5772\ldots$.

At the beginning of this Subsection we stated that it was 
sufficiently accurate to replace $R({\bf p}^2)$ by $c/2\mu$ in the 
series (\ref{U6}). Now it is necessary to prove that the 
correction to this approximation is of order $\alpha^7$ or 
greater. With the definition $\delta R=R({\bf p}^2)-c/2\mu$, the 
correction to the $\alpha^6$ term $IRIRI$ will be
\begin{equation}I\delta RI(c/2\mu )I+I(c/2\mu )I\delta RI\label{dcorr}\end{equation}
We must prove that this term is of order $\alpha^7$ only. Since 
$R_E-c/2\mu =\mbox{\rm O}(\alpha^2)$, we have $\delta R=\Delta R+\mbox{\rm O}
(\alpha^2)$, and it is 
sufficient to prove that
\begin{equation}I\Delta RI(c/2\mu )I+I(c/2\mu )I\Delta RI\label{DCORR}\end{equation}
is of order $\alpha^7$.

The calculations of Subsection C suggest that 
(\ref{DCORR}) should be of order $\alpha^7$, since $\left<I\Delta 
RI\right>$ was 
of order $\alpha^5$ while $\left<I(c/2\mu )I\right>$ is of order $
\alpha^4$.  However, 
it is not quite obvious, since $\Delta R(\infty )=-R_E\approx -c/
2\mu$, and 
therefore the rules of power-counting 
would permit (\ref{DCORR}) to be of order $\alpha^6$.  To see 
whether it is or not needs an explicit calculation.  

In momentum space, each term of (\ref{DCORR}) contains 
a factor of the form 
\begin{equation}\int\frac {d\,{\bf l}}{(2\pi )^3}\frac {e^2}{({\bf p}
-{\bf l})^2}\Delta R({\bf l}^2)\frac {e^2}{({\bf l}-{\bf q})^2}\label{NLOT}\end{equation}
(preceded or succeeded by other factors) while in the 
first ($\alpha^6$) term of (\ref{U60SERIES}) the corresponding 
factor is
\begin{equation}\int\frac {d\,{\bf l}}{(2\pi )^3}\frac {e^2}{({\bf p}
-{\bf l})^2}\frac c{2\mu}\frac {e^2}{({\bf l}-{\bf q})^2}\label{LOT}\end{equation}
We will exhibit counterpart terms in these two 
expressions and show that the one in (\ref{NLOT}) is 
smaller than the other by a factor O$(\alpha )$.

Again it is necessary to use the Schwinger unit sphere. 
With the mapping ${\bf p}\to\xi$ as before, as well as ${\bf l}\to
\eta$ and 
${\bf q}\to\zeta$, it is true that \cite{SCHW64} 
\begin{eqnarray}
\frac {e^2}{({\bf p}-{\bf l})^2}&=&\frac {1-\xi_4}{\beta}\frac {e^
2}{(\xi -\eta )^2}\frac {1-\eta_4}{\beta}\label{COUL3TO4}\\
\frac {e^2}{(\xi -\eta )^2}&=&(2\pi )^3\alpha\sum_{NLM}\frac 1NY_{
NLM}(\xi )Y^{*}_{NLM}(\eta )
\label{COUL4}\end{eqnarray}

Using the various relations given in Subsection C, 
expression (\ref{LOT}) becomes for fixed $L$,$M$ 
\begin{equation}\frac {1-\xi_4}{\beta}\sum_{N'N^{\prime\prime}}\frac {
(2\pi )^3\alpha^2}{N'N^{\prime\prime}}Y_{N'LM}(\xi )A^{(c/2\mu )}_{
L,N'N^{\prime\prime}}Y^{*}_{N^{\prime\prime}LM}(\zeta )\frac {1-\zeta_
4}{\beta}\label{LOT4}\end{equation}
in which
\begin{equation}A^{(c/2\mu )}_{L,N',N^{\prime\prime}}=\int d\Omega_{
\eta}Y^{*}_{N'LM}(\eta )\frac {\beta c}{2\mu}\frac 1{1-\eta_4}Y_{
N^{\prime\prime}LM}(\eta )\label{A0LNN}\end{equation}
With the notations $N'=n'+L+1$, $N^{\prime\prime}=n^{\prime\prime}
+L+1$, this is
\begin{equation}A^{(c/2\mu )}_{L,N',N^{\prime\prime}}=\left[2^{L+
1}\Gamma (L+1)\right]^2\sqrt {\frac {n'!N'}{2\pi\Gamma (N'+L+1)}}\sqrt {\frac {
n^{\prime\prime}!N^{\prime\prime}}{2\pi\Gamma (N^{\prime\prime}+L
+1)}}B^{(c/2\mu )}_{L,N',N^{\prime\prime}}\label{A0LLNB}\end{equation}
where, with $y\equiv\eta_4$,
\begin{equation}B^{(c/2\mu )}_{L,N',N^{\prime\prime}}=\frac {\beta 
c}{2\mu}\int^{+1}_{-1}dy\frac {(1-y^2)^{L+\frac 12}C^{L+1}_{n'}(y
)C^{L+1}_{n^{\prime\prime}}(y)}{1-y}\label{B0LNN}\end{equation}

Expression (\ref{B0LNN}) can be evaluated with the help 
of equation (\ref{z}). Supposing for definiteness that 
$n'>n^{\prime\prime}$, we easily find
\begin{equation}B^{(c/2\mu )}_{L,N',N^{\prime\prime}}=\frac {\beta 
c}{2\mu}\frac {\sqrt {\pi}\Gamma (L+\frac 12)\Gamma (N^{\prime\prime}
+L+1)}{n^{\prime\prime}!\Gamma (2L+2)\Gamma (L+1)}\label{B0LNNVAL}\end{equation}

Next we do the corresponding decomposition of 
(\ref{NLOT}). The steps are the same, with $\Delta R$ replacing 
$c/2\mu$. Using the representation (\ref{TOTDERIV}) as before,
we end up with 
\begin{equation}B^{(\Delta R)}_{L,N',N^{\prime\prime}}=-\frac 83\,
\delta_{L,0}\,\frac {\beta^2}{\mu^2}N'N^{\prime\prime}\left[\frac {
\mu^3}{m^3}+\frac {\mu^3}{M^3}\right]-\beta J_{L,n',n^{\prime\prime}}\label{BDRLnnVAL}\end{equation}
where
\begin{equation}J_{L,n',n^{\prime\prime}}=\int_{-1}^{+1}dyF(y)\frac 
d{dy}\left\{\sqrt {1+y}(1-y^2)^LC^{L+1}_{n'}(y)C^{L+1}_{n^{\prime
\prime}}(y)\right\}\label{JLnn}\end{equation}
The expression for $J_{L,n',n^{\prime\prime}}$ is evaluated the same way 
that (\ref{JLn}) was.  Similarly to $J_{L,n}$, it is found to be 
of order $\alpha^2$.

Therefore, recalling that $\beta$$=\mbox{\rm O}(\alpha )$, we finally have 
\begin{equation}B^{(\Delta R)}_{L,N',N^{\prime\prime}}=\mbox{\rm O}
(\alpha^2)+\mbox{\rm O}(\alpha^3)\label{BDR}\end{equation}
while
\begin{equation}B^{(c/2\mu )}_{L,N',N^{\prime\prime}}=\mbox{\rm O}
(\alpha )\label{B0}\end{equation}
Thus the expression (\ref{NLOT}) is of order $\alpha$ smaller 
than (\ref{LOT}), and so the 
correction term (\ref{DCORR}) is a factor $\alpha$ smaller than 
the original quantity $I(c/2\mu )I(c/2\mu )I$ which is of order 
$\alpha^6$. It follows that the sum (\ref{U60}) does correctly 
represent the series (\ref{U6}) up to and including order 
$\alpha^6$.

\subsection{Summary}

The complete energy corrections of order $\alpha^5$, $\alpha^6\log
(1/\alpha )$, 
and $\alpha^6$, to the Bethe-Salpeter equation (\ref{VFNBSE}) 
with a scalar Coulomb kernel, are given by the sum of:

\begin{itemize}
\item\  Equation (\ref{DELE6BS}), from the two-body 
Bohr-Sommerfeld formula (\ref{BOHRSOMM});
\item\   Equation (\ref{DELE6DELRVAL}), from the 
correction $I\Delta RI$ to $U$;
\item\   Equation (\ref{DELEU6L1}) for $L\ge 1$, and equation 
(\ref{DELEU6L0}) for $L=0$, which came from the sum
(\ref{U60}) of the rest of the Coulomb series for $U$.
\end{itemize}\

These corrections were calculated for every bound state, 
not just for a few low-lying states. The first correction was 
calculated by simple algebra. The latter two were 
calculated from first-order perturbation theory. As far 
as we know, calculations like these are beyond the power of 
the Salpeter reduction.

\section{Conclusion}

Advantages of the formalism shown here are that the 
local wave equation is derived directly from the 
Bethe-Salpeter equation, and that the two-body wave 
equation is accurate enough to predict the energy 
levels of the Bethe-Salpeter equation to first 
relativistic order correctly.

Two-body relativistic wave equations such as 
(\ref{COULWVEQN}) are easier to solve than Salpeter 
equations such as (\ref{SAL}), because they are local in 
configuration space.  The non-local components of the 
Salpeter equation, such as the free-particle 
kinetic-energy operator $\sqrt {{\bf p}^2+m^2}+\sqrt {{\bf p}^2+M^
2}$, and the 
correction factor $Z_p$ shown in equation (\ref{ZSAL}), are 
replaced in the two-body relativistic wave equation by 
constants, such as $E=\sqrt {m^2-\beta^2}+\sqrt {M^2-\beta^2}$, 
$Z_E=(m+M)/E$, and the correction factor $R_E$ shown in 
equation (\ref{RE}).

In addition, when the kernel of the scalar Bethe-Salpeter 
equation is a scalar Coulomb potential, the relativistic 
two-body bound-state wave equation is soluble exactly.  
The resultant two-body Bohr-Sommerfeld formula not 
only predicts the Bethe-Salpeter energy levels to order 
$\alpha^4$ correctly, but it also allows much of the $\mbox{\rm O}
(\alpha^6)$ 
correction to be evaluated by simple algebra.  
Furthermore, the remaining $\alpha^5$ and $\alpha^6$ corrections are 
calculated by first-order perturbation theory only.  No 
second-order perturbation calculations are needed.

The results shown above suggest that it may be easier 
to solve Bethe-Salpeter equations, at least those whose 
binding interaction is a Coulomb potential, with the aid 
of a local wave equation rather than a Salpeter equation.  
It is hoped that the formalism developed here can be 
adapted to real two-body atoms such as the hydrogen 
atom and positronium.

I thank the trustees of Springfield Technical Community 
College for a half-time sabbatical leave which assisted 
the completion of this work.  I am grateful to G. Adkins 
for extensive comments.

\end{document}